\documentclass[aps,pra,reprint,superscriptaddress,showpacs]{revtex4-1}
\usepackage{graphicx,amsmath,amssymb,ulem}
\usepackage{verbatim}
\usepackage{ragged2e}
\usepackage[usenames]{xcolor}
\usepackage{cleveref}

\begin{document}
\definecolor{Green}{rgb}{0,0.6,0}

\title{Projective filtering of a single spatial radiation eigenmode}
\author{A.~M.~P\'erez}
\affiliation{Max-Planck Institute for the Science of Light, \\ Guenther-Scharowsky-Str. 1 / Bau 24, Erlangen  D-91058, Germany}
\affiliation{University of Erlangen-N\"urnberg, Staudtstrasse 7/B2, 91058 Erlangen, Germany}
\author{P.~R.~Sharapova}
\affiliation{Physics Department, Moscow State University, Leninskiye Gory 1-2, Moscow 119991, Russia}
\author{S.~S.~Straupe}
\affiliation{Physics Department, Moscow State University, Leninskiye Gory 1-2, Moscow 119991, Russia}
\author{F.~M.~Miatto}
\affiliation{Department of Physics, University of Ottawa, Ottawa, ON, Canada K1N 6N5}
\author{O.~V.~Tikhonova}
\affiliation{Physics Department, Moscow State University,  Leninskiye Gory 1-2, Moscow 119991, Russia}
\affiliation{Skobeltsyn Institute of Nuclear Physics, Lomonosov Moscow State University, Moscow 119234, Russia}
\author{G.~Leuchs}
\affiliation{Max-Planck Institute for the Science of Light, \\ Guenther-Scharowsky-Str. 1 / Bau 24, Erlangen  D-91058, Germany}
\affiliation{University of Erlangen-N\"urnberg, Staudtstrasse 7/B2, 91058 Erlangen, Germany}
\author{M.~V.~Chekhova}
\affiliation{Max-Planck Institute for the Science of Light, \\  Guenther-Scharowsky-Str. 1 / Bau 24, Erlangen  D-91058, Germany}
\affiliation{University of Erlangen-N\"urnberg, Staudtstrasse 7/B2, 91058 Erlangen, Germany}
\affiliation{Physics Department, Moscow State University, Leninskiye Gory 1-2, Moscow 119991, Russia}

\begin{abstract}
Lossless filtering of a single coherent (Schmidt) mode from spatially multimode radiation is a problem crucial for optics in general and for quantum optics in particular. It becomes especially important in the case of nonclassical light that is fragile to optical losses. An example is   bright squeezed vacuum generated via high-gain parametric down conversion or four-wave mixing. Its highly multiphoton and multimode structure offers a huge increase in the information capacity provided that each mode can be addressed separately. However, the nonclassical signature of bright squeezed vacuum, photon-number correlations, are highly susceptible to losses. Here we demonstrate lossless filtering of a single spatial Schmidt mode by projecting the spatial spectrum of bright squeezed vacuum on the eigenmode of a single-mode fiber. Moreover, we show that the first Schmidt mode can be captured by simply maximizing the fiber-coupled intensity. Importantly, the projection operation does not affect the targeted mode and leaves it usable for further applications.
\end{abstract}
\maketitle

\section{Introduction}

Sources with a perfectly single-mode spatial spectrum are desirable but rare - an example is a laser with the beam quality factor  $M^2=1$. Most %of
sources contain multiple spatial modes, which leads to  the need for filtering methods. Preferably, these methods should be lossless, i.e., they should maintain all energy contained in the filtered mode. This is important for laser sources but becomes absolutely crucial for certain types of nonclassical light, because of the destructive role of losses.

Although formally one can choose free-space modes in many different ways, the most common example being plane-wave modes, spatial coherence dictates a special choice of eigenmodes for each type of radiation. For instance, if a mode has to contain all spatially coherent radiation, it has to be chosen according to the Mercer expansion~\cite{Mandel} of the first-order Glauber's correlation function. The Mercer expansion provides the so-called \textit{coherent modes} because the radiation within each of them is coherent.

Very similar to Mercer expansion is the Schmidt decomposition. While the Mercer expansion describes coherence of partially coherent light, the Schmidt decomposition also accounts for photon-number correlations. Within a certain Schmidt mode the radiation is coherent and has photon-number correlations only with itself or with a single matching mode. Such modes, which typically are not monochromatic plane waves, have been used to describe
nonclassical light, mostly for frequency/temporal modes and sometimes also for wavevector/spatial modes. Bennink and Boyd introduced the term `squeezing modes' for describing squeezing within a broad frequency spectrum of a traveling-wave parametric amplifier \cite{Boyd2002}. Opatrny and coworkers used the same concept to describe frequency correlations for Kerr-squeezed pulses in optical fibres~\cite{Opatrny}. It is worth stressing that the `squeezing modes' of Ref.~\cite{Boyd2002} as well as the `broadband modes' of Ref.~\cite{Opatrny} and the Schmidt modes mentioned further here are the same eigenmodes, the ones that diagonalize the Hamiltonian producing the radiation. It has been shown~\cite{Boyd2002} that without a proper selection of such modes, the degree of measured squeezing decreases considerably. This is a consequence of the fragility of squeezing to losses. In the case of multimode bright squeezed vacuum (BSV), generated through high-gain parametric down-conversion (PDC)~\cite{PDC} or four-wave mixing (FWM)~\cite{FWM}, the necessity to filter out a single Schmidt mode in a clean way, without losing its photons or admixing photons from different modes, is especially important due to strong thermal fluctuations within each mode~\cite{Rohde,Rytikov,Agafonov}. If unmatched modes are detected, these fluctuations completely destroy the squeezing.

There exists a partial solution in the case of homodyne detection, where one detects only the modes matching the ones of the local oscillator. It is therefore possible to select proper modes by tailoring the radiation of the local oscillator. In a series of works~\cite{Fabre}, Fabre, Treps and coworkers studied BSV generated by frequency combs and its eigenmodes (referred to there as `supermodes'). They detected the eigenmodes selectively using a specially tailored local oscillator. Similar strategy of tailoring the local oscillator is applied in experiments with spatially multimode BSV produced via FWM~\cite{Embrey}. However, in homodyne detection one cannot make use of the radiation in the selected mode, for instance, by coupling it with atoms or mechanical systems. If we seek any further use of nonclassical light, we need another type of projective filtering, a non-destructive one.

As a solution, here we consider filtering of a spatial mode with an optical fibre. We will show in Section~\ref{sec:projective} that an optical fibre performs the projection of the input spatial spectrum on its eigenmode, which can be considered, to a good approximation, as a Gaussian beam. We should stress that only in the case where the fibre eigenmode coincides with the radiation Schmidt mode, the filtering procedure will retain all features of the initial radiation such as coherence, peculiarities of the photon statistics, non-classicality, etc. We test the quality of such filtering by measuring the fraction of intensity transmitted through the fibre and comparing it  with the theoretical expectation, which is the weight of the strongest mode in the Schmidt decomposition. As the radiation to be tested, we choose multimode BSV generated through high-gain PDC.

It is worth mentioning that the quality of filtering a single coherent mode of PDC with an optical fibre has been never tested in experiment before. Many authors aimed at maximizing the total efficiency of coupling low-gain PDC (SPDC) radiation into an optical fibre~\cite{coupling}. It has been shown that the optimal case is the one of rather tightly focused pump, when SPDC contains few, almost a single, mode; but even under this condition the maximum coupling efficiency does not exceed $75\%$~\cite{coupling}. Besides, as we will show further, namely in this case the coupling of a single eigenmode is lossy. Other authors~\cite{Smirr,Dixon14,Guerreiro13} maximize the heralding efficiency of signal and idler SPDC photons coupled into single-mode fibers. It was found that the heralding efficiency is maximal and nearly $100\%$ for a softly focused pump and, correspondingly, for spatially multimode SPDC~\cite{Dixon14}. At the same time, to the best of our knowledge, no one considered the losses within a single eigenmode. In order to determine the shapes and weights of the eigenmodes for PDC radiation, we apply the Schmidt decomposition, which is considered in Section~\ref{sec:eigenmodes}. We describe the experiment in Section~\ref{sec:experiment} where we assess the quality of the filtering.

In the experiment, we deal with spatial (near-field) and angular (far-field) modes. Being parts of Fourier-conjugate spectra, they are alternative ways to describe the radiation. In Section~\ref{sec:frequency} we consider the analogy of projective filtering for temporal/frequency spectra and propose a method for linear projective filtering of these modes. Finally, in Section~\ref{sec:conclusion} we summarize the results.

\section{Projective filtering and spatial radiation eigenmodes}\label{sec:projective}
\subsection{Filtering with an aperture} If one needs to select a single mode from the angular (spatial) spectrum, the simplest strategy is to put an aperture of a certain size into the far (near) field. For a very small aperture, the mode selected this way will be a spherical wave in the first case and a plane wave in the second one. It is well known that the radiation after such filtering will be coherent (see, for instance, Ref.~\cite{Mandel}). As a consequence, the statistical properties (such as photon-number distribution) of certain radiation types will be also maintained. This will be the case, for instance, for thermal light~\cite{Mandel}. However, for more fragile types of light, e.g. squeezed vacuum, photon-number correlations will be lost~\cite{Rohde,Rytikov}. Moreover, even if the aperture has a specially chosen size and intensity transmission repeating the intensity distribution of a single mode, the filtering will still be lossy.

\subsection{Single-mode fibre as a projective filter} In order to maintain photon-number correlations, we require a different strategy, %the
one in which only a single eigenmode is filtered out, or a pair of conjugated eigenmodes, in the case of light with bipartite correlations. It is important for the filtering to be of projective type: a single field mode of the incident radiation should be projected on the eigenmode of the filtering device. This kind of filtering is provided by a single-mode fibre~\cite{Stas}, with the restriction that its eigenmode is a Bessel function, very close to a Gaussian. If a mode of any other shape has to be filtered, the fibre could be preceded by a spatial light modulator (SLM) performing %a
the transformation from this shape to a Gaussian~\cite{Stas,Miatto2}. The only losses introduced this way will be the ones associated with the SLM or any alternative device.

\begin{figure}[t]
\begin{center}
\includegraphics[width=0.38\textwidth]{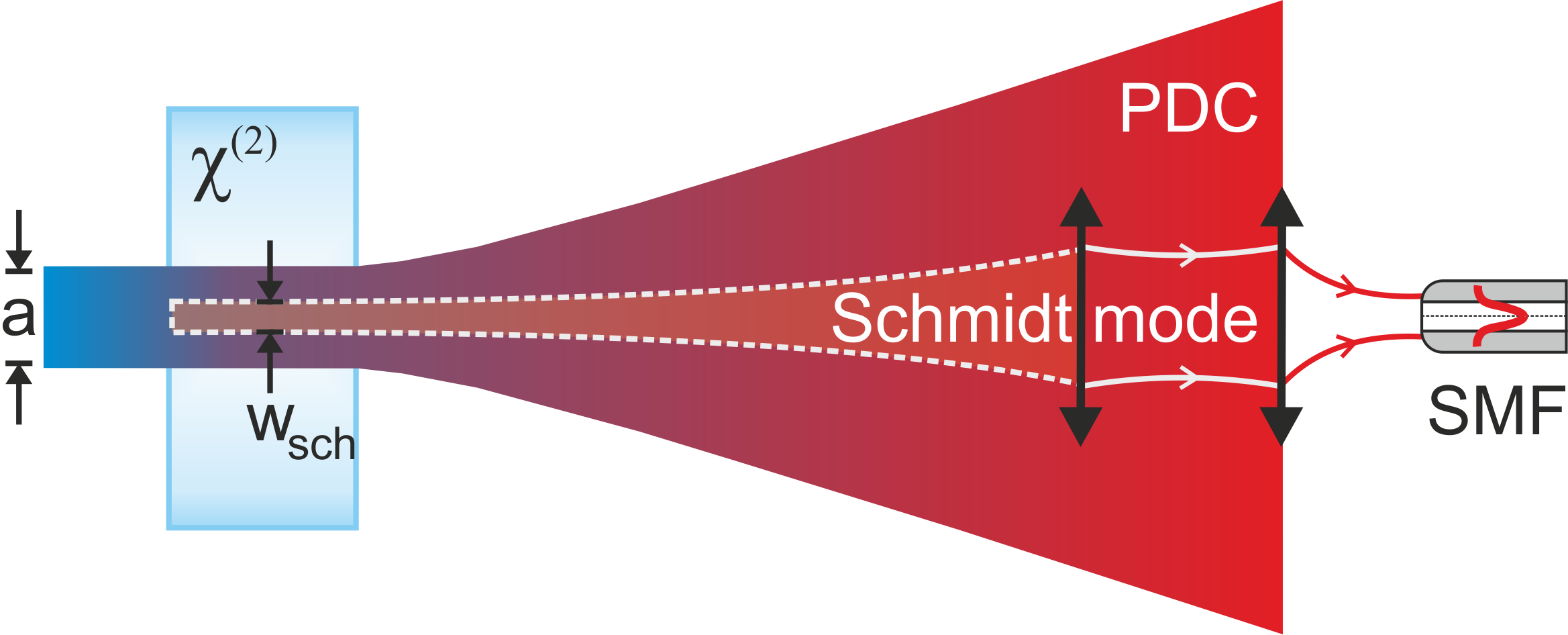}
\end{center}
\caption{Filtering a single eigenmode of broadband PDC radiation with a single-mode fibre.}
\label{fig:schemeschmidt}
\end{figure}

As mentioned before, here we consider PDC as a source of multimode radiation
(Fig.~\ref{fig:schemeschmidt}). The PDC radiation created in a nonlinear crystal
has its near-field effective diameter related to the full width at half maximum (FWHM) $a$ of the Gaussian pump. However, as it is multimode, its angular divergence is much larger than expected for a Gaussian beam of waist $a$. A single-mode fibre filters out the Gaussian Schmidt mode with the waist $w_{sch}$ losslessly and blocks all other modes.

Mathematically, the effect of the fibre is described as a projection operation. Let the eigenmode of the fibre in the near field be $f(\mathbf{r})$, $\int|f(\mathbf{r})|^2 d\mathbf{r}=1$, with $\mathbf{r}$ being the coordinate at the input facet of the fibre, and the radiation eigenmodes be $u_n(\mathbf{r})$, $\int u^*_n(\mathbf{r})u_k(\mathbf{r})d\mathbf{r}=\delta_{nk}$. Then we can write the photon annihilation operator $\hat{A}$ in the fibre mode as a linear combination of the photon annihilation operators $\hat{A}_k$ acting in the radiation eigenmodes, which form an orthonormal set:
\begin{equation}
\hat{A} = \sum_{k} C_k \hat{A}_k.
\label{eq:expansion}
\end{equation}
Here,
\begin{equation}
C_k \equiv\int f(\mathbf{r})u^*_k(\mathbf{r})d\mathbf{r}, \,\,\sum_k |C_k|^2=1,
\label{eq:projection}
\end{equation}
are the projections of the fibre mode on the radiation eigenmodes.
If a single radiation eigenmode $u_0(\mathbf{r})$ coincides with the fibre mode, its projection $C_0=1$ %,
  and the other projections are zero.

Equivalently, the modes can be described in the far field as $u_k(\mathbf{q})$, by introducing the transverse wavevector $\mathbf{q}$.
Note that the same relation (\ref{eq:expansion}) will be valid for classical fields instead of the annihilation operators.

\begin{figure}%[htb]
\begin{center}
\includegraphics[width=0.27\textwidth]{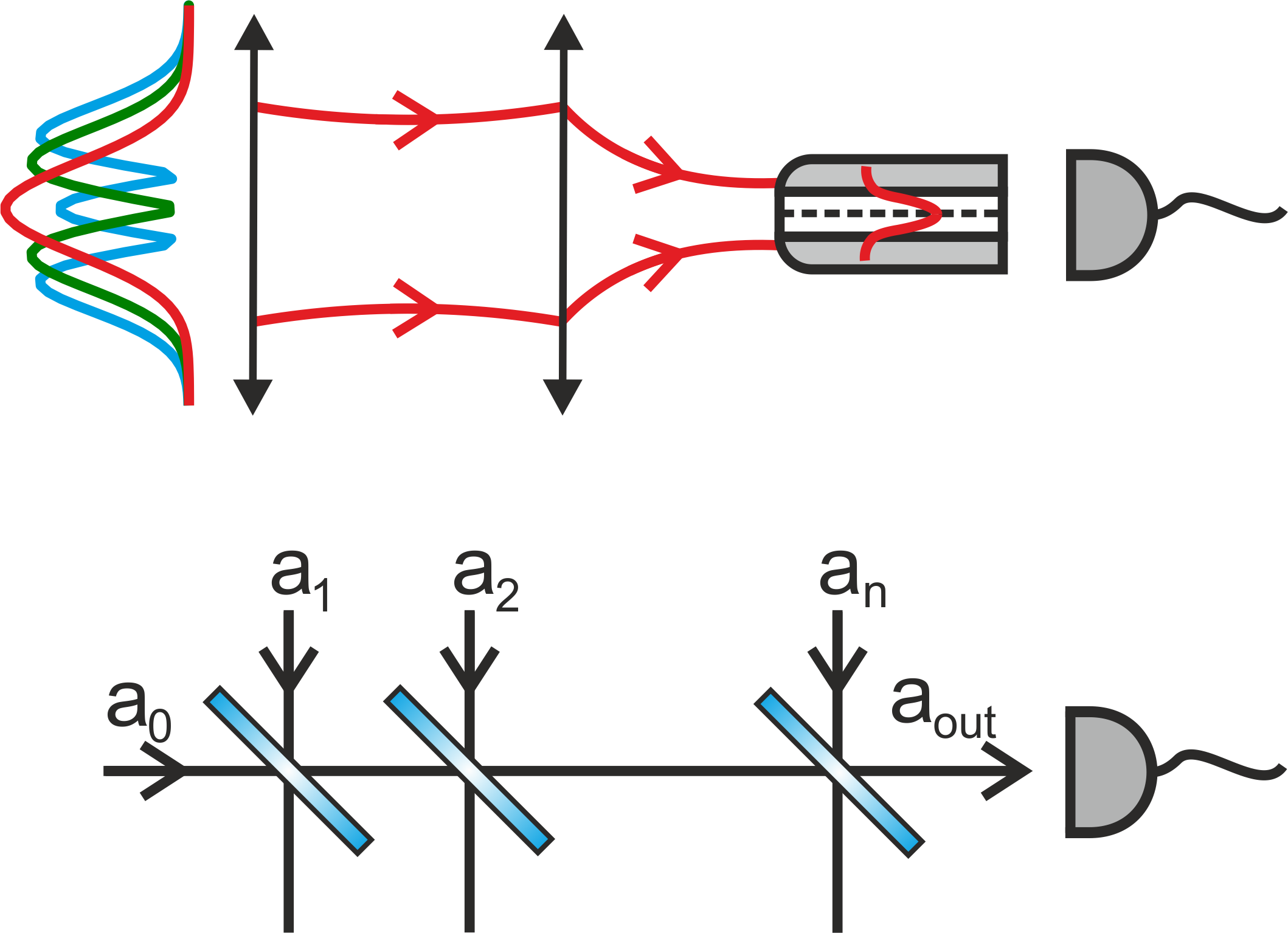}
\end{center}
\caption{Analogy between multiple orthogonal modes projected on the eigenmode of a fiber and multiple input modes of several beamsplitters projected on a single output mode.}
\label{fig:BS}
\end{figure}

Relation (\ref{eq:expansion}) is the same as the one describing the field or operator at one output of a series of beamsplitters in terms of the fields (operators) at their inputs (Fig.~\ref{fig:BS}):
 \begin{equation}
a_{out} = \sum_{k=0}^n C_k a_k,\,\,\sum_{k=0}^n|C_k|^2=1.
\label{eq:BS}
\end{equation}
 The projections $C_k$ depend then on the field transmission/reflection coefficients $t_k,r_k$ :
 \begin{equation}
C_0=t_1t_2\dots t_n,\,C_1=r_1t_2\dots t_n,\dots,\,C_n=r_n.
\label{eq:BS}
\end{equation}

\subsection{Spatial radiation modes} In the classical case, the eigenmodes of free-space radiation are given by the Mercer expansion, defined through the first-order Glauber's correlation function $G^{(1)}(\mathbf{r}_1,\mathbf{r}_2)$~\cite{Mandel,Bobrov},
\begin{equation}
G^{(1)}(\mathbf{r}_1,\mathbf{r}_2) = \sum_{n} \alpha_{n}u^*_{n}(\mathbf{r}_1)u_{n}(\mathbf{r}_2).
\label{eq:Mercer}
\end{equation}
The modes $u_n(\mathbf{r})$ are called coherent modes because the radiation within each such a mode is coherent.

Whenever photon correlations between two beams are of interest, especially nonclassical photon-number correlations, rather than coherence within a single beam, Mercer decomposition is not sufficient and the Schmidt decomposition should be used, as will be shown in the next section addressing high-gain PDC.

\section{High-gain PDC and its eigenmodes}\label{sec:eigenmodes}
High-gain PDC is a convenient way to produce bright squeezed vacuum, a macroscopic quantum state of light that is among the most promising sources for quantum technologies. It manifests polarization and photon-number entanglement~\cite{review,ent} and can violate Bell's inequality under certain experimental conditions~\cite{Bell}. It already found applications in quantum imaging~\cite{imaging} and quantum metrology~\cite{metrology}, in particular enabling phase super-sensitivity~\cite{supersensitivity}.
BSV is multimode in angle and frequency and can have the mean photon number per mode as high as $10^{13}$~\cite{single-mode}. These features provide its high information capacity as quantum information can be encoded in the photon number of each mode.
Ideally one would like to isolate and efficiently control
each mode without losing its nonclassical correlations.

\subsection{The Schmidt decomposition and the Bloch-Messiah reduction} The eigenmodes of BSV are found from the Schmidt decomposition, in which each mode of the signal beam is correlated to a
single idler mode~\cite{Law,Fedorov09,Stas,Christ,Eckstein,Sharapova14}. In signal and idler channels taken separately, the modes found this way coincide with the coherent modes from the Mercer expansion (\ref{eq:Mercer})~\cite{Miatto1,Bobrov}.

It has been shown~\cite{Sharapova14} that BSV exhibits the same Schmidt modes as two-photon light generated via low-gain PDC under the same
experimental geometry. The modes are found by diagonalizing the Hamiltonian of PDC,
\begin{equation}
\hat H =i\hbar\Gamma \int d\mathbf{q}_s d\mathbf{q}_i F(\mathbf{q}_s,\mathbf{q}_i)\hat a^{\dagger}_{\mathbf{q}_s}\hat a^{\dagger}_{\mathbf{q}_i} + h.c.,
\label{eq:hamiltonian}
\end{equation}
where $\Gamma$ is the coupling parameter scaling as the pump amplitude, $\mathbf{q}_s,\mathbf{q}_i$ the signal and idler transverse wavevector components, $\hat a^{\dagger}_{\mathbf{q}_s}\hat a^{\dagger}_{\mathbf{q}_i}$ the photon creation operators in the corresponding plane-wave modes, and $F(\mathbf{q}_s,\mathbf{q}_i)$ the two-photon amplitude (TPA). This term, as it is used here, is conventional: at strong pumping, photons are no more emitted in pairs but in large even numbers. The Hamiltonian is diagonalized through the Schmidt decomposition of the TPA,
\begin{equation}
F(\mathbf{q}_s,\mathbf{q}_i) = \sum_{m,n} \sqrt{\lambda_{mn}}u_{mn}(\mathbf{q}_s)v_{mn}(\mathbf{q}_i),
\label{eq:decomposition}
\end{equation}
where $\lambda_{mn}$ are the Schmidt eigenvalues, $\sum_{m,n}\lambda_{mn}=1$, and $u_{mn}(\mathbf{q}_s),v_{mn}(\mathbf{q}_i)$ are the 2D Schmidt modes of the signal and idler radiation. In the degenerate case, where signal and idler beams are indistinguishable, $u_{mn}(\mathbf{q})=v_{mn}(\mathbf{q})$.

After substituting Eq.~(\ref{eq:decomposition}) into Eq.~(\ref{eq:hamiltonian}), the Hamiltonian can be written in the diagonal form (the Bloch-Messiah reduction)~\cite{Sharapova14},
\begin{equation}
    H=i\hbar \Gamma\sum\limits_{k}\sqrt{\lambda_{k}}\left(\hat{A}^\dagger_{k} \hat{B}^\dagger_{k} - \hat{A}_{k} \hat{B}_{k}\right),
\label{eq:Hamiltonian_Schmidt}
\end{equation}
where $k\equiv\{m,n\}$ and $\hat{A}^\dagger_{k}$, $\hat{B}^\dagger_{k}$ are photon creation operators for the Schmidt modes defined as
\begin{equation}
    \begin{split}
        &\hat{A}^\dagger_{k} = \int d\mathbf{q}_s u_{k}(\mathbf{q}_s)a^\dagger_{\mathbf{q}_s},\\
        &\hat{B}^\dagger_{k} = \int d\mathbf{q}_i v_{k}(\mathbf{q}_i)a^\dagger_{\mathbf{q}_i},\\
    \end{split}
\label{eq:Schmidt_CA}
\end{equation}
In order to label the photons as signal/idler, we assume here some additional degree of freedom, for instance, different wavelengths as in non-degenerate PDC.  The solution to the Heisenberg equations for the operators in Schmidt modes leads to Bogolyubov-type transformations between the input operators $\hat{A}_{k0},\hat{B}_{k0}$ and the output ones $\hat{A}_k,\hat{B}_k$~\cite{Sharapova14},
\begin{equation}
    \begin{split}
        \hat{A}_k =         &c_k\hat{A}_{k0} + s_k \hat{B}^\dagger_{k0}, \\
        \hat{B}_k =         &c_k\hat{B}_{k0} + s_k\hat{A}^\dagger_{k0},
    \end{split}
\label{eq:Bogolyubov}
\end{equation}
where
\begin{equation}
c_k =\cosh\left(G\sqrt{\lambda_k}\right), \,\,s_k =\sinh\left(G\sqrt{\lambda_k}\right),
\end{equation}
and $G=\int \Gamma dt$ is the parametric gain.

The mean photon number in mode $k$ is $N_k=s_k^2$, and the total mean photon number in each (signal/idler) beam is given by the incoherent sum over separate Schmidt modes: $N=\sum_{k}s_k^2$. Thus, the Schmidt eigenvalues at high gain are renormalized~\cite{Sharapova14},
\begin{equation}
\lambda'_{k}=\frac{s_k^2}{N}=\frac{\sinh^2\left(G\sqrt{\lambda_k}\right)}{\sum_{k}\sinh^2\left(G\sqrt{\lambda_k}\right)},
\label{eq:renormalization}
\end{equation}
so that the eigenvalue $\lambda'_{k}$ of a certain Schmidt mode determines the contribution of this mode to the total number of photons,  $N_k=N\lambda'_{k}$.
The effective number of Schmidt modes, in the low-gain regime given by the Schmidt number $K=[\sum_{k} \lambda_{k}^2]^{-1}$, is hence reduced at high gain,
$K'=[\sum_{k}\lambda_{k}'^{2}]^{-1}$.

The signal photon annihilation operator $\hat{A}$ after the fibre is then given by Eq.~(\ref{eq:expansion}), with $\hat{A}_k$ being the photon annihilation operators in the signal Schmidt modes and the projections $C_k$ given by Eq.~(\ref{eq:projection}). A similar expression is valid for the idler photon annihilation operator $\hat{B}$ after the fibre: $\hat{B} = \sum_{k} D_k \hat{B}_k$, with $D_k$ denoting the projections of the idler Schmidt modes on the fibre eigenmode.
We assume here that the fiber mode function is identical for the signal and idler photons. The mean photon numbers after the fibre are
\begin{equation}
    \begin{split}
        &N_{s}  = \langle \hat{A}^\dagger \hat{A}\rangle = \sum\limits_k  \left|C_k\right|^2 s_k^2=N\sum\limits_k  \left|C_k\right|^2 \lambda'_k,\\
        & N_{i}  = \langle \hat{B}^\dagger \hat{B}\rangle = \sum\limits_k  \left|D_k\right|^2 s_k^2=N\sum\limits_k  \left|D_k\right|^2 \lambda'_k.
    \end{split}
    \label{eq:phnumbers}
\end{equation}

\subsection{The two-crystal scheme} In our experiment, described further in Sec.~\ref{sec:experiment}, PDC is generated in two consecutive nonlinear crystals placed into a common Gaussian pump beam. For two crystals of length $L$ at a distance $l$, the normalized TPA for frequency-degenerate or nearly degenerate type-I PDC has the form~\cite{Klyshko,Sharapova14}

\begin{equation} \label{eq:TPA}
\begin{aligned}
F(\mathbf{q}_s,\mathbf{q}_i)&=\exp\left[-\frac{a^2 k_0^2}{8\ln2}\lbrace \left(\theta_s^2+\theta_i^2\right)+2\theta_s\theta_i \cos(\phi_s-\phi_i)\rbrace \right] \times \\
&\times \text{sinc}\left(\frac{\Delta k_z L}{2}\right)\cos\left(\frac{\Delta k_z L + \Delta k'_z l }{2}\right) \times \\
&\times \exp(-i \Delta k_z L) \exp \left(\frac{-i \Delta k'_z l}{2}\right),
\end{aligned}
\end{equation}
where $k_0$ is the length of signal and idler wavevectors, $\theta_{s,i},\phi_{s,i}$ are their spherical angles,
$a$ is the full width at half maximum (FWHM) of the pump intensity distribution,
$\Delta k_z=k_p - k_0[\cos(\theta_s)+\cos(\theta_i)]$ is the longitudinal mismatch inside each crystal and $k_p$ is the length of the pump wavevector. In its turn, $\Delta k'_z=k_p^{air} - k_0^{air}[\cos(\Theta_s)+\cos(\Theta_i)]$
is the longitudinal mismatch in the air gap between the crystals, where the pump and signal/idler wavevectors take values $k_p^{air}$ and $k_0^{air}$, respectively; $\Theta_{s,i}=\frac{n_0}{n_{0}^{air}}\theta_{s,i}$, and $n_{0}$, $n_{0}^{air}$ are the signal/idler refractive indices inside the crystals and inside the air gap, respectively.

The Schmidt decomposition is most conveniently found in the cylindrical frame of reference, in which the transverse wavevectors $\mathbf{q}_{s,i}$ are given by their modules, $q_{s,i}=k_0\sin\theta_{s,i}$, and the azimuthal angles, $\phi_{s,i}$~\cite{Miatto1,Miatto2}. In this case,
$F(\mathbf{q}_s,\mathbf{q}_i)$ can be written as a Fourier expansion due to its periodicity in $(\phi_s-\phi_i)$~\cite{Miatto2},
\begin{equation}
F(q_s,q_i,\phi_s-\phi_i)=\sum_n \chi_n(q_s,q_i) e^{i n (\phi_s-\phi_i)},
\label{TPA polar}
\end{equation}
where $\chi_n(q_s,q_i)$ can be found using the inverse Fourier transformation. Then, the Schmidt decomposition of $\chi_n(q_s,q_i)$ yields
\begin{equation}
\chi_n(q_s,q_i)=\sum_m \sqrt{\lambda_{mn}}\frac{\tilde{u}_{mn}(q_s)}{\sqrt{q_s}} \frac{\tilde{v}_{mn}(q_i)}{\sqrt{q_i}},
\label{decomposition}
\end{equation}
with the functions $\tilde{u}_{mn}(q_s)$ and $\tilde{v}_{mn}(q_i)$ obeying the normalization condition,
\begin{equation}
\int_0^{\infty} d q_s \tilde{u}_{mn}(q_s)\tilde{u}^{*}_{kn}(q_s)=\int_0^{\infty} d q_i \tilde{v}_{mn}(q_i)\tilde{v}^{*}_{kn}(q_i)=\delta_{mk}.\nonumber
\label{orthogonalisation}
\end{equation}
From Eq.~(\ref{decomposition}), we obtain the Schmidt decomposition of the TPA (\ref{eq:decomposition})
 with $u_{mn}(\mathbf{q}_{s})=\frac{\tilde{u}_{mn}(q_s)}{\sqrt{q_s}}e^{i n \phi_s}$, $v_{mn}(\mathbf{q}_{i})=\frac{\tilde{v}_{mn}(q_i)}{\sqrt{q_i}}e^{-i n \phi_i}$.

\subsection{Filtering the first Schmidt mode} The first Schmidt mode $u_{00}$ of the BSV state, a Gaussian of waist
$w_{sch}$, can be filtered by projecting the angular spectrum on the
eigenmode of a fiber, which is close to a Gaussian of waist $w$,  $f(\mathbf{q})=(\sqrt{\pi} w)^{-1} \exp (-q^2/2w^2)$.
The projections of different Schmidt modes on the eigenmode of the fibre are (see Eq.~\ref{eq:projection})
\begin{equation}
C_{mn}=\int_0^{2\pi}d\phi \int_0^\infty qdq f(q,\phi) u_{mn}(q, \phi),
\label{eq:projection_coeffs}
\end{equation}
$\sum_{m,n} |C_{mn}|^2 = 1$.
Note that since the fiber mode does not depend on $\phi$, only Schmidt modes with $n=0$ will contribute in the coupling efficiency.
%In particular~\cite{Qassim},
%\begin{align}
%|C_{00}^w|^2=\frac{4}{2+\frac{w^2}{w_{sch}^2}+\frac{w_{sch}^2}{w^2}}.
%\label{eq:Gaussian}
%\end{align}

From (\ref{eq:phnumbers}), the coupling efficiency $T$ for the signal radiation is obtained by summing the photon numbers
that couple into the fibre from different modes:
\begin{equation}
T=\sum_{m,n} |C_{mn}|^2 \lambda'_{mn}\le\lambda'_{00}.
\label{eq:coupling}
\end{equation}
The last inequality follows from the fact that $\lambda'_{m\ge0,n\ge0}\le\lambda'_{00}$ and shows that the photon number transmitted through the fibre is maximized when the fibre mode exactly matches the first Schmidt mode,
$|C_{00}|=1$, and $T_{max}=\lambda'_{00}$. In this case the first Schmidt mode is filtered perfectly, without losses. Only in this case the filtered Schmidt mode maintains all specific features of the incoming PDC radiation including non-classicality.

\subsection{The absence of losses} However, for the first Schmidt mode to coincide with the fibre eigenmode, the former should be real or at least have no spatially varying phase, as the latter is a real Gaussian function. At the same time, Eq.~(\ref{eq:TPA}) for the TPA includes some phase factors; moreover, even the TPA for a single crystal is complex~\cite{Sharapova14}. As a consequence the Schmidt modes of down-converted radiation can be complex functions. In this case the projection amplitudes $C_{mn}$ entering Eq.~(5) of the main text are reduced. Using the Cauchy-Schwarz inequality one can show that the maximal projection amplitude for the first Schmidt mode $u_{00}(q, \phi)$ and the fiber eigenmode $f(q,\phi)$ will be still less than 1 if the Schmidt mode is complex. Indeed,
\begin{eqnarray}
|C_{00}|^2 \leq \int_0^{2\pi}d\phi \int_0^\infty q dq |f(q,\phi)|^2 \times\nonumber \\
\times\int_0^{2\pi}d\phi \int_0^\infty q dq |u_{00}(q, \phi)|^2 = 1,
\label{eq:projection_coeffs}
\end{eqnarray}
with the equality valid only in the case where the Schmidt mode and the fiber eigenmode coincide up to a constant phase factor. Whenever $|C_{00}|<1$, the filtering procedure leads to intrinsic losses.
\begin{figure}[htb]
\begin{center}
\includegraphics[width=0.4\textwidth]{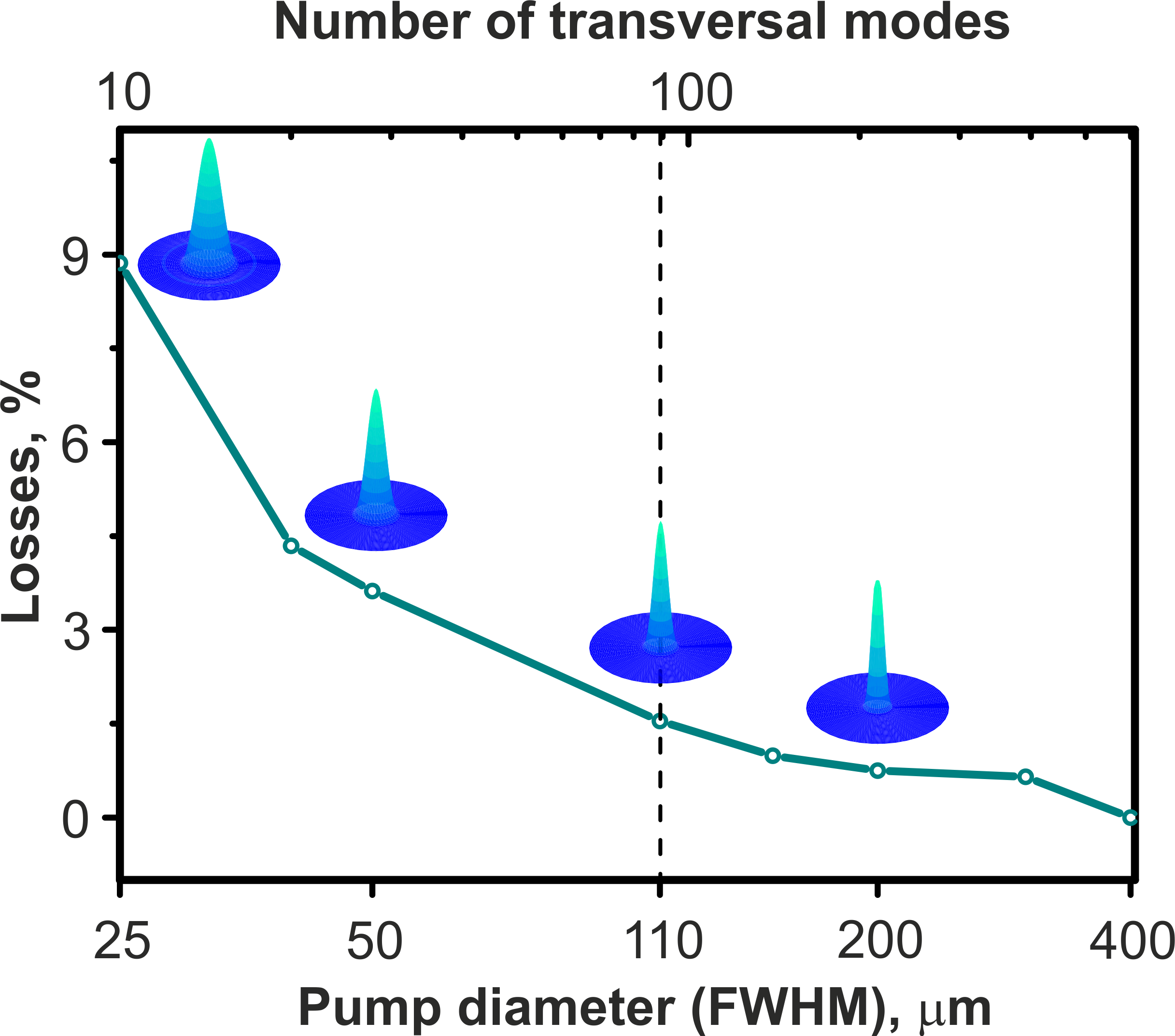}
\end{center}
\caption{Numerical estimation of losses accompanying the filtering of PDC generated in the two-crystal system. The pump diameter $a$ is varied in order to change the initial number $K$ of the transversal modes. The dashed line corresponds to the parameters used in our experiment.
The insets show the intensity distributions of the first Schmidt mode for various points in the transverse-wavevector space.}
\label{fig:loss}
\end{figure}
Now let us consider under what circumstances the first Schmidt mode would be strongly complex. The TPA~(\ref{eq:TPA}) is complex due to the phase factor $\exp\{-i\Delta k_z L-i\Delta k'_z l/2\}$, which does not depend on the pump beam waist $a$. At the same time, the size of the first Schmidt mode does depend on $a$. By choosing appropriate $a$, one can find the conditions under which the phase of the TPA does not vary noticeably within the scale of the first Schmidt mode. In this case, the left part of (\ref{eq:projection_coeffs}) will be close to $1$ and the losses will be negligibly small.

To demonstrate this, we have performed the Schmidt decomposition and calculated $C_{00}$ for various diameters $a$ of the pump. For each case, the shape of the first Schmidt mode was calculated (the corresponding intensity distributions in transverse wavevector space are shown as insets in Fig.~\ref{fig:loss}). Further, the losses accompanying the filtering, given by $1-|C_{00}|^2$, were numerically calculated (shown in Fig.~\ref{fig:loss} as points connected by a line.) The result is that the weaker the focusing, the smaller the losses. %(
From the inset, one can also notice that for weaker focusing, the first Schmidt mode is narrower and this is why the TPA phase in its vicinity is flat.
%)
For focusing into more than $100$ $\mu$m, the losses are less than $1.5\%$. In our experiment, we use $a=110$ $\mu$m, which corresponds to $2\%$ of losses (shown in Fig.~\ref{fig:loss} by dashed line).

Thus, for a weakly focused pump, the first Schmidt mode has a flat phase and can be filtered nearly losslessly using a single-mode fibre. In the opposite case of a tightly focused pump, projective filtering with a single-mode fibre is lossy. In principle, even in this case, the phase of the Schmidt mode to be filtered can be eliminated before coupling the light into the fibre, but this requires some special efforts like using an SLM.

Note that experiments with SPDC report high (up to $96\%$) heralding efficiencies with a softly focused pump~\cite{Dixon14, Guerreiro13}, despite a low generation rate of photons. This is in agreement with our observation (Fig.~\ref{fig:loss}): if the heralded mode coincides with the Schmidt mode of a highly multimode state, the losses of its fibre coupling can be indeed negligible.
%Our work may contribute to the interpretation of such results: the high multimodeness of such unfiltered emission makes the phase variation on each individual mode smaller.  The heralding-efficiency in this kind of systems would be large if a single coherent mode of the radiation is losslessly filtered either through one or two channels -depending on the system configuration-, a task that becomes possible when the phase of the mode to be filtered is closer to a flat phase.
In contrast, with the pump tightly focused into the crystal, so that the resulting number of modes is small~\cite{coupling}, the coupling efficiency is never high, again in agreement with our calculations (Fig.~\ref{fig:loss}). At the same time, as we have mentioned above, the efficiency of filtering a single mode has been never directly measured before. The next section considers this measurement.
\section{Experiment}\label{sec:experiment}
\subsection{Intensity measurements}
\begin{figure}[htb]
\begin{center}
\includegraphics[width=0.40\textwidth]{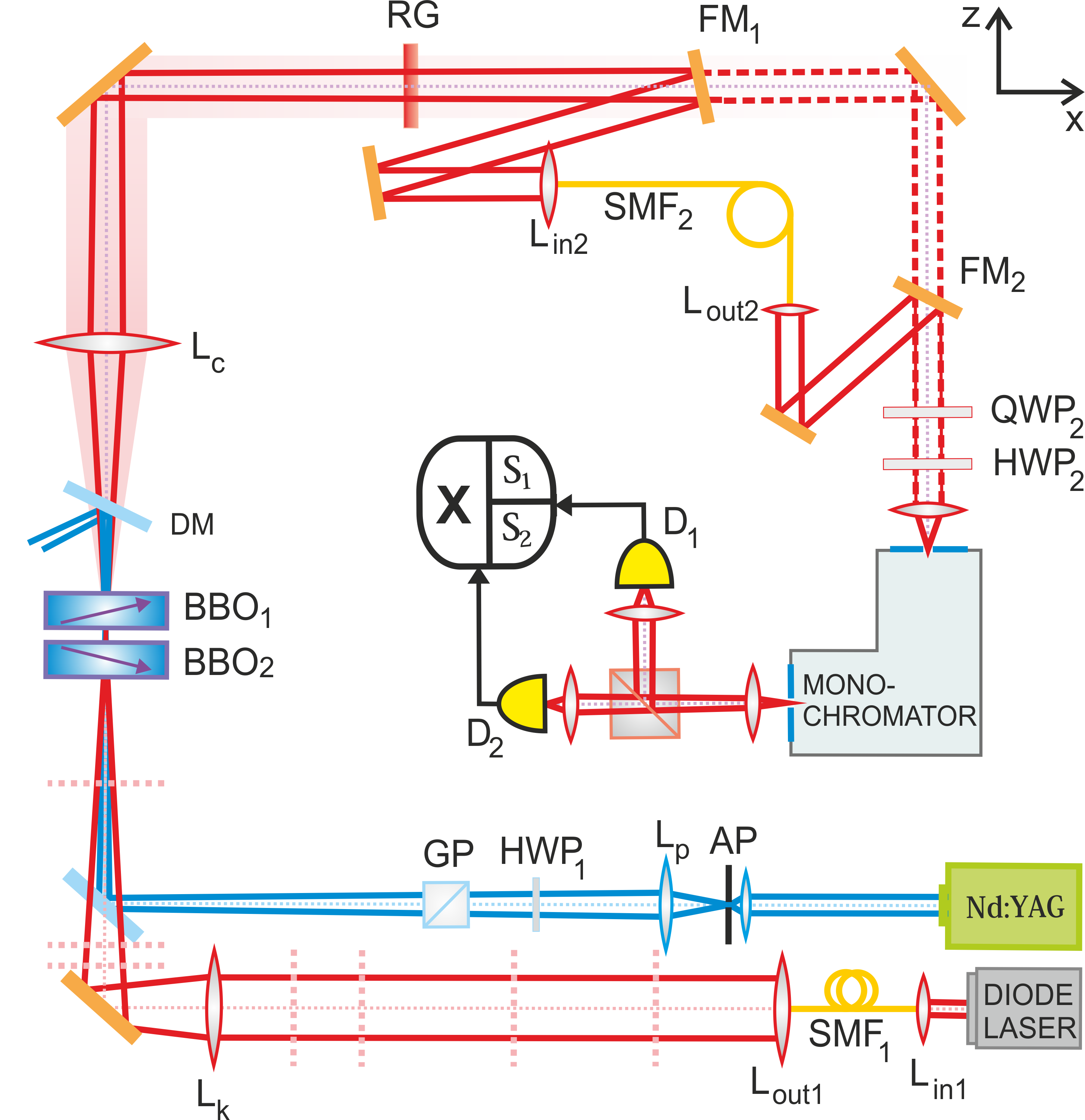}
\end{center}
\caption{Setup scheme. The third harmonic of a Nd:YAG laser, mode cleaned and properly polarized (aperture AP,
half-wave plate HWP$_1$
and Glan prism GP), focused into crystals BBO$_1$ and BBO$_2$ by lens L$_p$, generates PDC.
An auxiliary laser diode beam, mode cleaned by a single-mode fiber SMF$_1$, serves to emulate the spatial distribution of different Gaussian modes, with the waists in the same plane. These modes are prepared with the help of different
lenses L$_k$, placed at the planes marked by the vertical dotted lines one at a time. A Gaussian mode is filtered by the
single-mode fiber SMF$_2$ after lenses L$_c$,L$_{in2}$. Flipping mirrors FM$_1$ and FM$_2$ separate the filtering path
from the free-space path (dashed lines). Before PDC enters the monochromator, its polarization is adjusted by half-wave and quarter-wave plates (HWP$_2$,QWP$_2$).
Finally, the signals are analyzed in a Hanbury Brown-Twiss interferometer including detectors $D_1$ and $D_2$.}
\label{fig:setup}
\end{figure}

\begin{figure}%[htb]
\begin{center}
\includegraphics[width=0.47\textwidth]{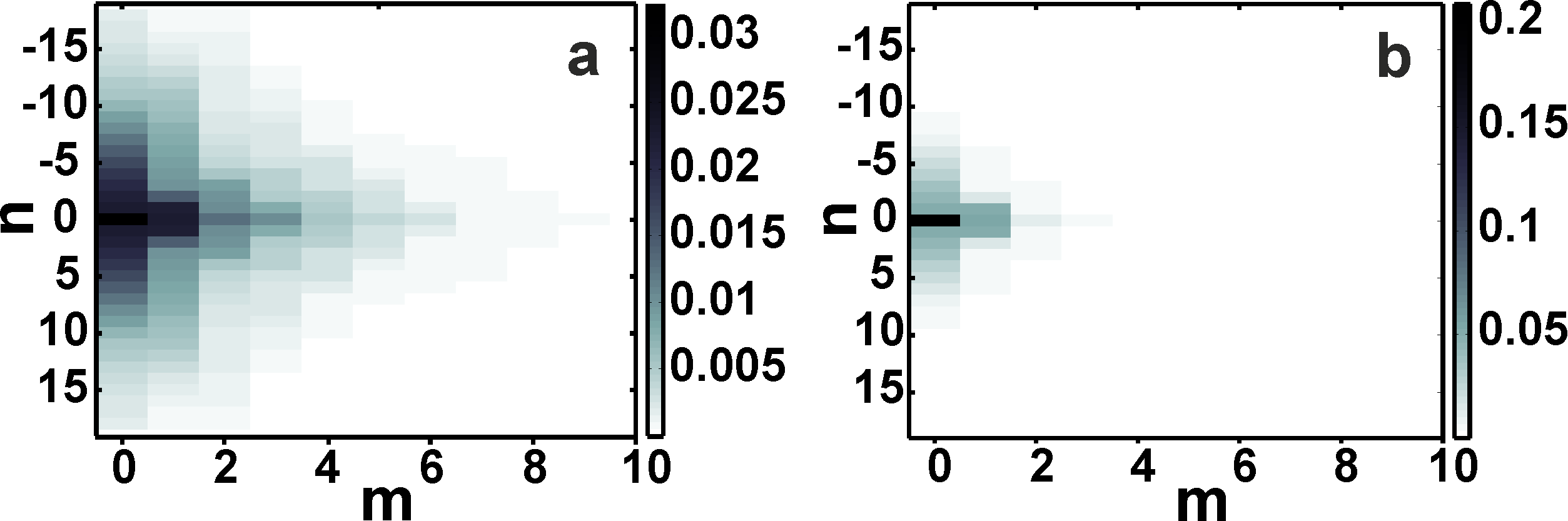}
\end{center}
\caption{The Schmidt eigenvalues for PDC from the two-crystal scheme used in our experiment (a)
in the low-gain regime, resulting in the
effective mode number $K=89$, and (b) in the high-gain regime, resulting in $K'=14.7$.}
\label{fig:comparison}
\end{figure}

The  BSV state is generated via high-gain PDC (Fig.~\ref{fig:setup}) in two 1 mm long beta barium borate crystals (BBO$_1$, BBO$_2$) cut for type-I
collinear degenerate phase matching and arranged in the anisotropy compensating configuration at the closest achievable distance of $2.5\pm 0.5$ mm~\cite{anisotropy1,anisotropy2,separation}.
The crystals are pumped by a Nd:YAG laser third harmonic (wavelength $355$ nm, pulse duration 18 ps, and repetition rate 1 kHz). The pump is mode-cleaned by means of a diamond pinhole
(AP) and its polarization is set by a Glan prism (GP)
and a half-wave plate HWP$_1$. The pump waist is imaged by lens $L_p$ (100 mm focal length) onto the plane between the crystals,
where its FWHM is 110 um $\pm$ 5 um.
This experimental configuration in the low-gain regime creates a two-photon state with the spatial Schmidt number $K=89$
(Fig.~\ref{fig:comparison}a). The first Schmidt mode is very close to a Gaussian function, with the divergence $0.006$ rad and the waist $53$ $\mu$m. Our experiment is performed in the high-gain regime (pump power $70$ mW), which, according to Eq.~(\ref{eq:renormalization}), leads to the
reduction in the number of spatial modes. The resulting Schmidt number is calculated
to be $K'=14.7$ (Fig.~\ref{fig:comparison}b) for $G=22.8$.

After the crystals,
the pump radiation is cut off by a dichroic mirror (DM) and a red glass filter RG 650 (RG).
The spatial filtering is performed by a standard step index single-mode fiber (SMF$_2$, Thorlabs SM600).
In the experiment we filter from the PDC radiation, with the help of the fibre, Gaussian beams with different waists/divergences and measure the fraction $T$ of PDC intensity transmitted through the fibre. According to Eq.~(\ref{eq:coupling}), the largest $T$ should equal the first Schmidt eigenvalue and should be achieved when the coupled Gaussian mode coincides with the first Schmidt mode.

 For emulating Gaussian beams with various waists, we use an additional CW diode laser with the wavelength $706.5$ nm.  The diode laser beam is mode-cleaned by a single-mode fiber SMF$_1$, collimated by lens L$_{out1}$, and then overlapped with the pump beam on the crystals after passing through one of the lenses L$_k$,
with the other lenses removed from the beam. Each lens is aligned so that
it forms the beam waist of a certain diameter $w_k$ (measured by a beam profiler) on the crystals. For each waist diameter, the beam is
coupled into SMF$_2$ by means of lens L$_c$ ($150$ mm focal length) and aspheric lens L$_{in2}$ ($3.3$ mm focal length),
with the losses $12-18\%$ including $4$\% reflection at the uncoated input facet. For each lens L$_k$, efficient coupling of the diode laser beam into the fiber indicates that the system filters out a Gaussian mode with a certain waist $w_k$ and the corresponding divergence $\Delta\theta_k$. After this, the diode laser is switched off and the spatial filtering is applied to the PDC radiation. Light out-coupled from the fiber is sent through a monochromator  selecting a
bandwidth of $0.1$ nm around the non-degenerate wavelength $708$ nm. Zero-order quarter-wave plate (QWP$_2$) and half-wave plate (HWP$_2$) optimize the incoming polarization on the monochromator and minimize losses.

For comparing the data in the presence and in the absence of filtering, a free-space channel is used where the
PDC radiation is sent to the monochromator directly, avoiding the fiber (Fig.~\ref{fig:setup}). Switching between the free-space and spatially
filtered channels is done with the flipping mirrors FM1 and FM2. The efficiency $T$ of coupling into the fiber is measured  by dividing the sum signal of the detectors $D_1$ and $D_2$
in the presence of filtering by the one in its absence.

\begin{figure}%[htb]
\begin{center}
\includegraphics[width=0.47\textwidth]{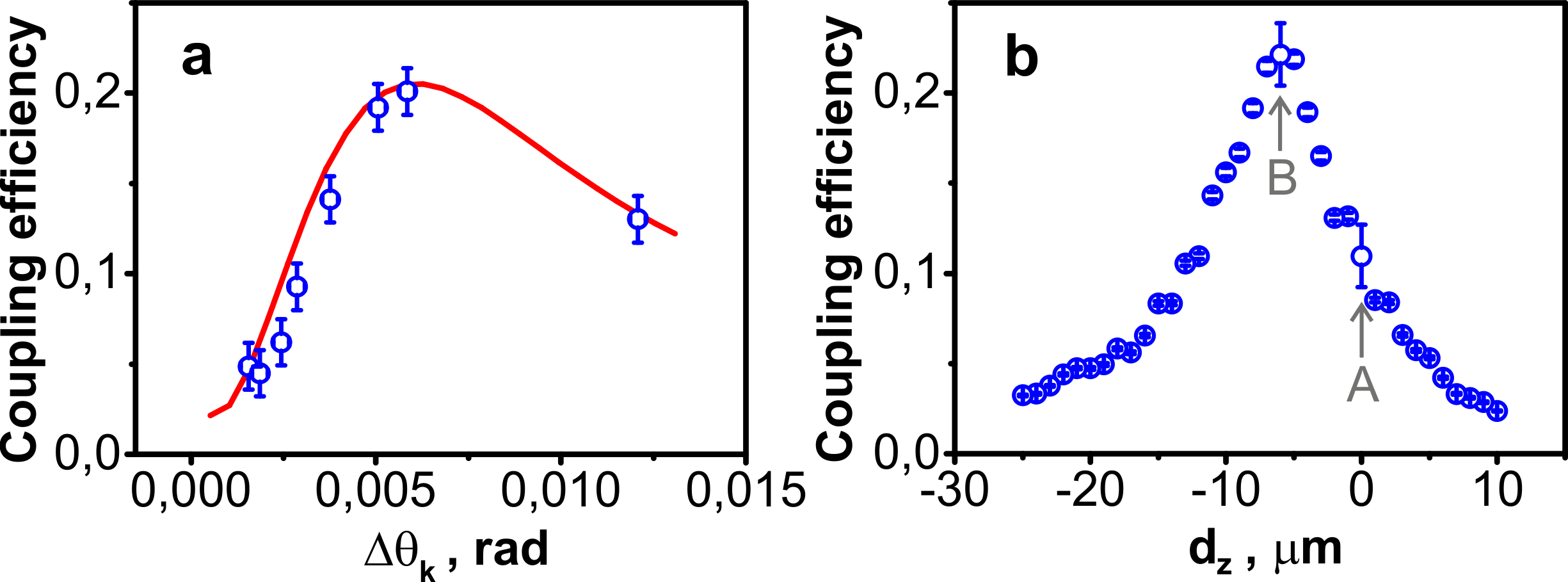}
\end{center}
\caption{Spatial filtering of PDC radiation: (a) the coupling efficiency versus the angular width $\Delta \theta_k$ of the Gaussian beam coupled into the fiber and (b) versus the longitudinal adjustment $d_z$ of the coupling system. The bars at points A and B show the absolute uncertainty of the coupling efficiency.}
\label{fig:fit}
\end{figure}

Since each Gaussian beam is coupled to SMF$_2$
with $85 \pm 3\%$ coupling efficiency, the PDC coupling is underestimated.
We quantify these technical losses and correct for them at each measured point.
In the free-space configuration we also measure the total number of spatial modes
through the normalized second-order intensity correlation function $g^{(2)}$ at the non-degenerate wavelength of $708$ nm, making sure that all angular spectrum is collected. Then, the measured correlation function depends on the number of modes $K'$ as $g^{(2)}=1+1/K'$~\cite{separation}.
We obtain $g^{(2)}=1.07 \pm 0.01$, which
indicates $K'=15 \pm 2$ spatial modes, in agreement with the calculation (Fig.~\ref{fig:comparison}b).

Fig.~\ref{fig:fit}a shows the coupling efficiencies for each angular width $\Delta\theta_k$ of the filtered Gaussian mode as well as the
theoretical curve, with a good agreement between the two.
Both curves have their maximum at an angular width of $6$ mrad, which is the angular divergence of the first Schmidt mode. Moreover, the maximum coupling efficiency achieved experimentally is $0.201\pm0.005$,
which matches the first Schmidt eigenvalue $\lambda'_{00}=0.205$ calculated for our BSV state (Fig.~\ref{fig:comparison}). From these values,
the losses of the filtering procedure do not exceed $2\%$, as expected from the calculation (Fig.~\ref{fig:loss}). This proves that filtering of the first Schmidt mode of
PDC radiation with a single-mode fiber is nearly lossless, up to reflections at the fiber facets and imperfect coupling.

Thus, the efficiency of coupling PDC radiation into the fiber is maximal when
the first Schmidt mode
coincides with the fiber eigenmode. One can guess that the first Schmidt mode can be targeted by simply maximizing the coupling efficiency.
In what follows, we show that this is indeed the case. In fact, this is the strategy usually applied for low-gain PDC~\cite{coupling} but up to now it has not been tested for the filtering of a single mode.

Starting from a setting where a `wrong' Gaussian mode is coupled into the fiber (first point from the right in Fig.~\ref{fig:fit}a and point A in Fig.~\ref{fig:fit}b), we improve the PDC coupling by varying the distance between lens L$_{in2}$ and the tip of the fiber.
This way, we are able to achieve the coupling efficiency equal to the first Schmidt eigenvalue (point B in Fig.~\ref{fig:fit}b). This indicates that the mode collected is indeed the targeted Schmidt mode. Note that we achieve this goal by simply moving the fibre tip. At first sight impossible, it is feasible in our scheme due to the fact that in Gaussian optics, the position of the waist image depends on the initial waist size. As a result, modification of the diode laser beam waist on the crystals mainly leads to the displacement of the beam waist after lens $L_{in2}$ rather than to the change of its size.

\subsection{Correlation measurements}
\begin{figure}%[htb]
\begin{center}
\includegraphics[width=0.3\textwidth]{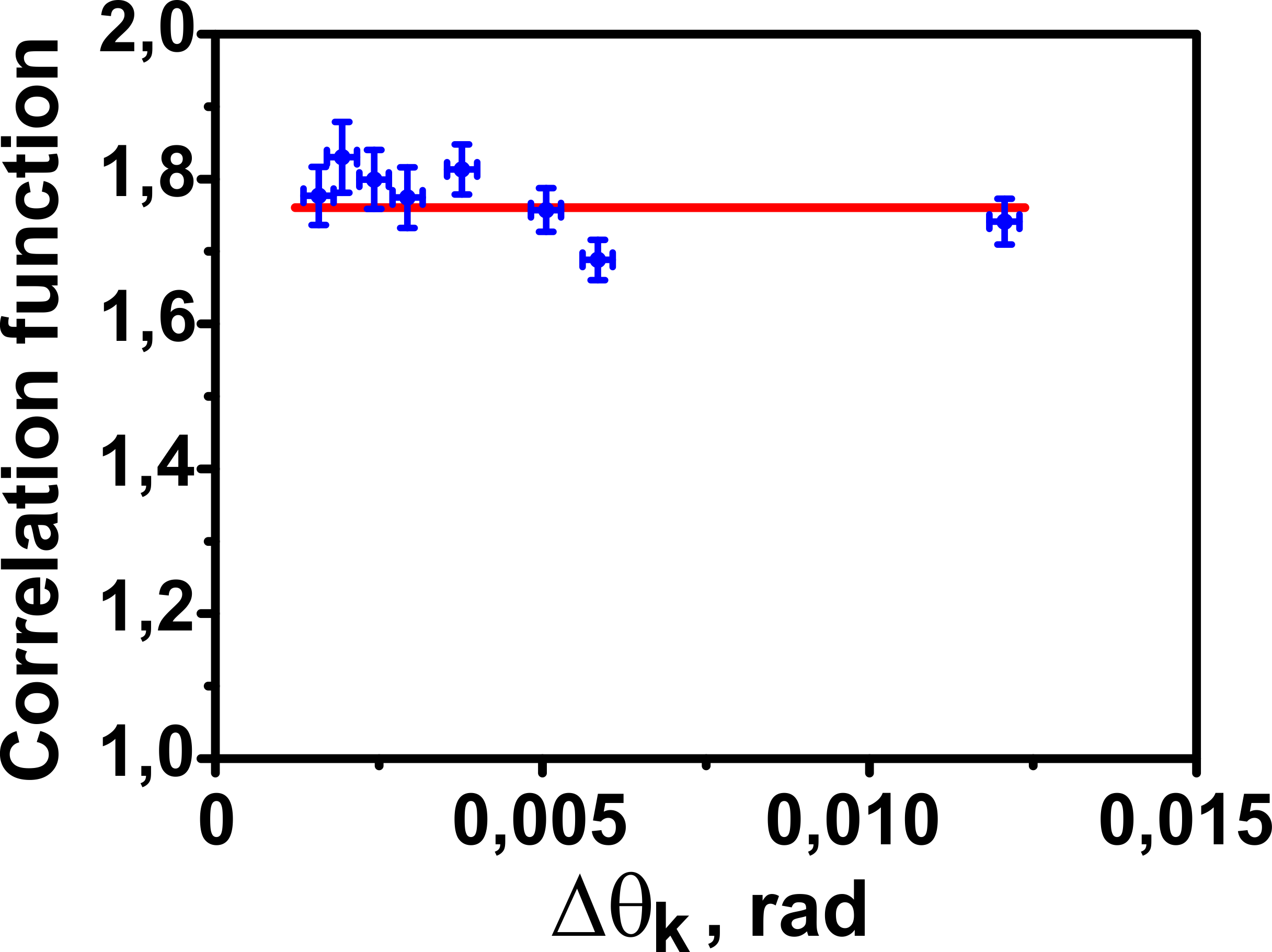}
\end{center}
\caption {$g_{ss}^{(2)}$ measurement at the non-degenerate wavelength $\lambda_{ndeg}=708$ nm versus the angular width of the Gaussian mode coupled into the fiber.}
\label{fig:g2ndeg}
\end{figure}

One might think that the mode filtering quality could be controlled by means of the correlation function measurement~\cite{Eckstein,Christ,separation}
since it usually indicates the number of modes.
However, the measured autocorrelation function of the signal radiation at zero time delay, $g_{ss}^{(2)}(0)$  (further, simply $g^{(2)}_{ss}$, Fig.~\ref{fig:g2ndeg}), is independent of the angular width of the Gaussian mode fully coupled into the fiber, and hence of the number of Schmidt modes contributing to the fiber output.

This can be understood from the analogy with a multi-port beamsplitter  (Fig.~\ref{fig:BS}). Indeed, with the thermal light at the input ports of a beamsplitter, the statistics of light at its each output port will be also thermal~\cite{Meda, Illusionist}.
Since each mode of PDC radiation has thermal statistics, the
radiation at the output of the fiber will be thermal, and one will measure $g_{ss}^{(2)}=2$ regardless of the number of modes contributing.
Since our frequency filtering is not perfect, the measured value is lower, $g_{ss}^{(2)}=1.76 \pm$ 0.02, as shown
in Fig. \ref{fig:g2ndeg}.

This shows that the quality of filtering cannot be assessed from the correlation function measurement for thermally populated Schmidt modes. At the same time, if the modes are populated with two-photon light, correlation measurements will indicate whether one or several modes are contributing to the fibre output.

Indeed, consider the normalized second-order correlation function at zero time delay after the spatial mode filter, the \textit{auto-correlation function}
\begin{equation}
    \begin{split}
        g^{(2)}_{ss} = &\frac{\langle A^\dagger A^\dagger A A \rangle }{ N_{s}^2} = \\
        & \frac{1}{N_{s}^2} \sum\limits_{i,j,k,l}(C_i)^* (C_j)^* C_k C_l \langle A_i^\dagger A_j^\dagger A_k A_l \rangle,
    \end{split}
\label{eq:auto-corr}
\end{equation}
and the \textit{cross-correlation function}
\begin{equation}
    \begin{split}
        g^{(2)}_{si} = & \frac{\langle A^\dagger B^\dagger A B \rangle}{N_{s}N_{i}} = \\
        & \frac{1}{N_{s}N_{i}} \sum\limits_{i,j,k,l}(C_i)^* (D_j)^* C_k D_l \langle A_i^\dagger B_j^\dagger A_k B_l \rangle.
    \end{split}
\label{eq:cross-corr}
\end{equation}
A straightforward calculation using (\ref{eq:Bogolyubov}) gives
\begin{equation}
    \begin{split}
        &\langle A_i^\dagger A_j^\dagger A_k A_l \rangle = s_i s_j s_k s_l (\delta_{jk}\delta_{il} + \delta_{ik}\delta_{jl}) \\
        &\langle A_i^\dagger B_j^\dagger A_k B_l \rangle = s_i c_j c_k s_l \delta_{ij}\delta_{kl} + s_i s_j s_k s_l \delta_{ik}\delta_{jl}.
    \end{split}
\label{eq:correlators}
\end{equation}
Substituting (\ref{eq:correlators}) into (\ref{eq:auto-corr}) and (\ref{eq:cross-corr}) we get the following expressions for the second-order correlation functions:
\begin{equation}
    \begin{split}
        &g^{(2)}_{ss} \equiv 2,\\
        &g^{(2)}_{si} = 1+ \frac{\left|\sum\limits_i C_i^* D^*_i s_i c_i \right|^2}{\left( \sum_i |C_i|^2 s_i^2 \right)\left( \sum_k |D_k|^2 s_k^2 \right)}.
    \end{split}
\end{equation}
The auto-correlation function is identically equal to $2$, in accordance with the multiport analogy and the results of our measurement (Fig~\ref{fig:g2ndeg}). The behavior of the cross-correlation function is more subtle. Its value depends on both coupling coefficients between the Schmidt modes and the fiber mode and the partial photon numbers in each Schmidt mode. In the simplest case where the fiber matches the first Schmidt mode, i.e. $\left|C_k\right| = \left|D_k\right| = \delta_{k0}$, the expression for $g^{(2)}_{si}$ simplifies to
\begin{equation}
    g^{(2)}_{si} = 2 +  \frac{1}{s_0^2} = 2 + \frac{1}{N_0},
    \label{g2(N)}
\end{equation}
where $N_0$ is the mean photon number in the first Schmidt mode. In this case the correlation function after the mode filter correctly describes the statistics of a single mode of the initial PDC source. However, when the mode-matching is not perfect and several modes have significantly non-zero coupling coefficients, the correlation function value deviates from the expression expected for a single-mode PDC field.
\begin{figure}[htb]
    \begin{center}
    \includegraphics[width=0.35\textwidth]{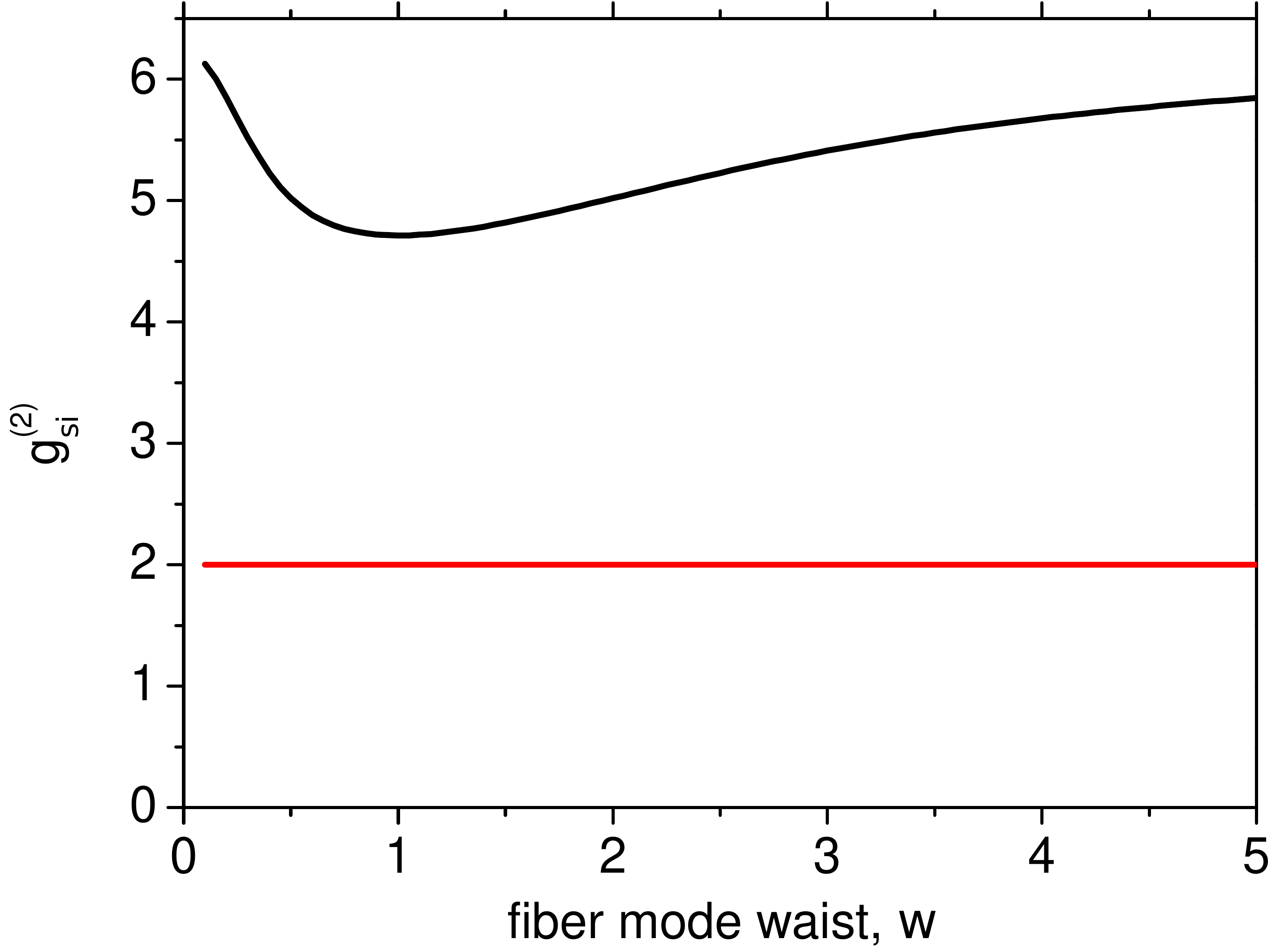}
\end{center}
\caption{Dependence of the cross-correlation function $g^{(2)}_{si}$ on the size of the mode coupled into the fiber for two values of the gain: $G=1$ (black line), and $G=10$ (red line). The width is given in relative units $w/w_0$, where $w_0$ is the width of the (approximately) Gaussian first Schmidt mode. In both cases the Schmidt number $K'=5$.}
\label{fig:g2_th}
\end{figure}
The dependence on the fiber mode waist, calculated using the double-Gauss model for the TPA, is shown in Fig.~\ref{fig:g2_th}. At first sight, it looks unusual that the normalized cross-correlation function is minimal for the case of optimal coupling. However, this is in line with the usual dependence of normalized cross-correlation function for PDC on the mean photon number, given by Eq.~(\ref{g2(N)}). At larger photon numbers, PDC always has lower normalized cross-correlation function. Therefore, optimal coupling, giving the maximal mean photon number, at the same time results in a minimum of the cross-correlation function.
Unfortunately, the dependence flattens out with increasing gain, making the effect unobservable under current experimental conditions.

\section{Temporal/frequency analogy}\label{sec:frequency}
A natural question arises whether a similar technique can be applied to filtering a single frequency mode from a multimode spectrum. There have been proposals of doing this by means of up-conversion, with the pump mode properly tailored in frequency~\cite{Eckstein,up_theory}. A `quantum phase gate' based on this principle has been recently demonstrated to provide $80\%$ efficiency for coherent pulses at the input~\cite{up_experiment}. This kind of filtering is also of projective type as it projects the field mode on the eigenmode of the converter. However, the method is technically complicated, requires phase matching within a broad frequency range, may have additional losses for external radiation fed into the nonlinear converter and will be also influenced by noise whenever weak input radiation is considered (a feature of all similar up-converting devices).

\begin{figure}[htb]
    \begin{center}
    \includegraphics[width=0.4\textwidth]{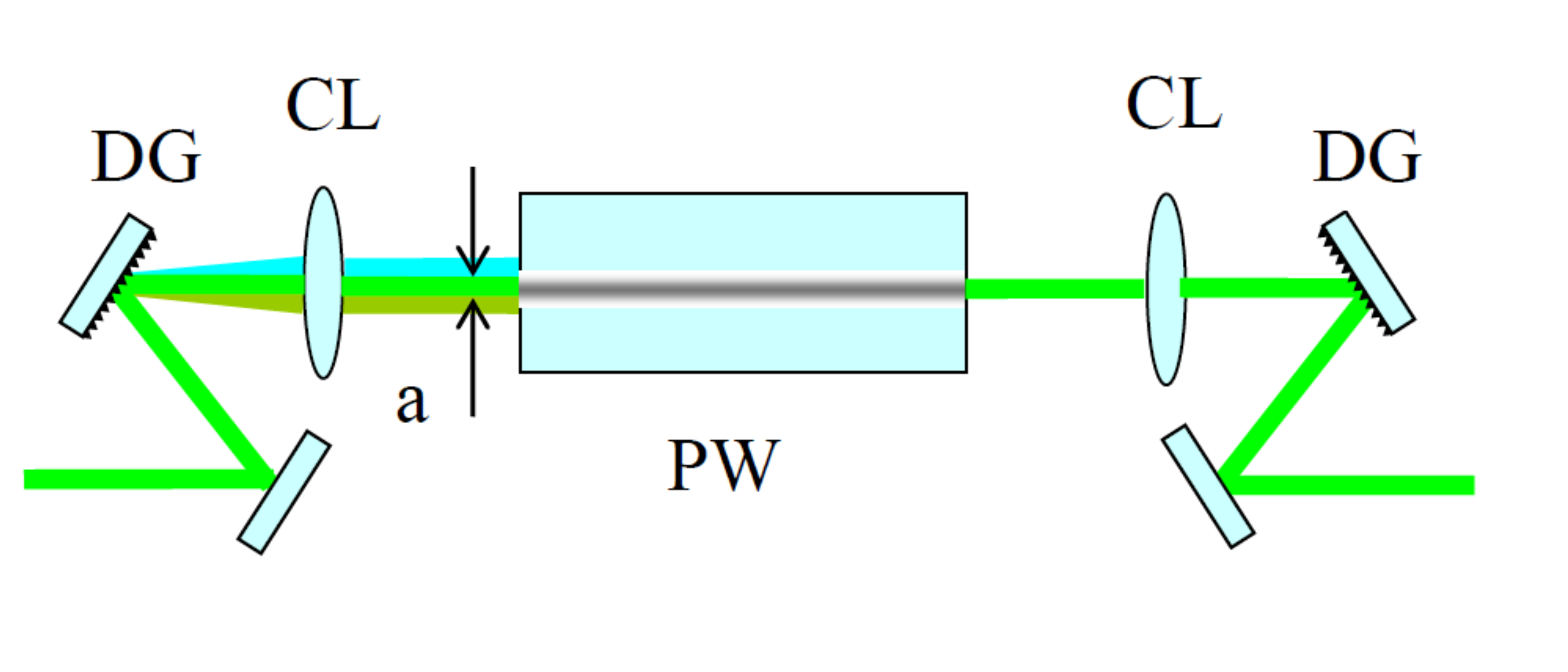}
\end{center}
\caption{A setup for linear projective filtering of the frequency modes. DG, diffraction grating; CL, cylindrical lens, PW, planar waveguide; $a$, the size of its eigenmode.}
\label{fig:freq}
\end{figure}

By analogy with the spatial filtering introduced above, we can suggest a similar scheme for the filtering of the frequency/temporal modes, based on converting the frequency into the angle and then filtering the angle. The proposed setup (Fig.~\ref{fig:freq}) is based on a 4f pulse shaper, in which the input multimode pulse falls on a diffraction grating followed by a cylindrical lens. In the focal plane of the lens the vertical coordinate scales linearly with the wavelength. In a usual 4f system, a spatially-selective device such as a slit or an SLM is placed, after which another cylindrical lens and a diffraction grating collect the radiation into a single beam. To make the filtering projective, one can replace the spatial filter in the middle of the 4f scheme by a planar waveguide whose eigenmode has a Gaussian profile in the vertical direction. The size $a$ of the mode should correspond to the frequency width of the Gaussian mode to be filtered. This device will project the frequency spectrum on a Gaussian one. If a more complicated frequency mode has to be filtered out, a mode converter (for instance, an SLM) should be placed before the planar waveguide.

\section{Conclusion}\label{sec:conclusion}
In this paper, we have considered the spatial filtering performed by a single-mode fibre and have shown that, in contrast to %a
the filtering performed by %with
an aperture, it is of projective type and therefore imposes no losses on the mode filtered out. An analogy between a fiber and a multiport beamsplitter has been drawn. Based on this analogy, we have
considered this type of filtering for the spatial modes of high-gain PDC. It was shown that only under the condition that the pump is focused into the crystal not too tightly, this filtering can be practically lossless. Otherwise, if the pump waist is small, the Schmidt modes have spatially non-uniform phases and the filtering will be lossy unless a phase correction is applied.

Further, we have demonstrated projective filtering of the first Schmidt mode from the spectrum of high-gain PDC in experiment. The total losses accompanying the filtering are only caused by imperfect alignment as well as the reflection on the uncoated fibre facet and do not exceed 15\%. Importantly, the radiation in the mode filtered out this way is not destroyed and is available for further use. The method can be extended to higher-order spatial modes by using an appropriate spatial-mode transformations,
for instance with the help of a spatial light modulator. Furthermore, it has been shown that the correct Schmidt mode can be filtered simply by maximizing the coupling into the fiber, provided that the apertures of the lenses do not clip significant portions of the radiation. The transmissivity of the fibre being close to the first Schmidt eigenvalue can be therefore used as a criterion for the selection of the first Schmidt mode. On the other hand, the autocorrelation function cannot be used for this purpose as it is independent on the number of modes contributing to the fibre output intensity. The cross-correlation function can be used to characterize the number of modes contributing but only at low parametric gain.

Finally, a similar technique has been proposed for the filtering of a single frequency mode out of the PDC spectrum. It is based on a standard 4f pulse-shaping scheme where a planar waveguide is used as a spatially selective element.

The research leading to these results has received funding from the EU FP7 under grant agreement No.
308803 (project BRISQ2). We also acknowledge partial financial support of the Russian Foundation for Basic Research, grants 14-02-31084 mol$_{-}$a and 14-02-00389$_{-}$a. P.R.Sh acknowledges support of the `Dynasty' foundation.

\end{document}